\documentstyle[12pt]{article}

\begin{document}
\newcommand{\be}{\begin{equation}}
\newcommand{\ee}{\end{equation}}
\newcommand{\al}{\alpha}

\begin{center}
{\large \bf Comment on\\
``Next-to-next-to-leading order vacuum polarization function of heavy quark
near threshold and sum rules for  
$b \bar b$ system'' \\and\\ 
``Next-to-next-to-leading order relation 
between $R(e^+e^-\rightarrow b\bar b)$
and $\Gamma_{\rm sl}(b\rightarrow cl\nu_l)$ 
and precise determination of $|V_{cb}|$''}\\[5mm]
  {\bf A.A.Penin and A.A.Pivovarov}\\[2mm]
  {\small {\em Institute for Nuclear Research of the
  Russian Academy of Sciences,}}\\
  {\small {\em 60th October Anniversary
  Pr., 7a, Moscow 117312, Russia}}\\[5mm]
\end{center}

\begin{center}
{\small \bf Abstract}
\end{center}
\vspace{-4mm}
{\small
The most recent 
recalculation 
of the two-loop correction to the static quark-antiquark
potential gave the numerical value different from the previously
known one.
We comment on the effect this change produces on the numerical 
estimates of the bottom quark pole mass $m_b$, 
the strong coupling constant 
$\al_s$ and the Cabibbo-Kobayashi-Maskawa matrix element
$|V_{cb}|$ obtained in our papers \cite{PP,PP1}.
}
\newpage
In two recent papers \cite{PP,PP1} 
numerical values of the bottom quark pole mass $m_b$, 
the strong coupling constant 
$\al_s$ and the Cabibbo-Kobayashi-Maskawa matrix element
$|V_{cb}|$ have been
determined
from the sum rules for the $\Upsilon$ system
and the $B$-meson 
semileptonic width. These phenomenological results
have been obtained by exploiting 
the next-to-next-to-leading 
order expression for the vacuum polarization function 
of a heavy quark near the
threshold. This expression 
depends on the value 
of the two-loop correction to the static quark-antiquark
potential.
In the analyses of refs.~\cite{PP,PP1} the numerical 
value of coefficient $a_2$
obtained in \cite{Peter} was used.
Recently the two-loop correction to the static potential has been
recalculated in \cite{YS} with a new result for the coefficient $a_2$
that differs from the previous one.
However we found that the use of corrected
numerical value of the coefficient $a_2$ leads
to a change of the numerical estimates
for $m_b$, $\al_s$ and $|V_{cb}|$ obtained in our papers
that lies well within the error bars given for these parameters 
in \cite{PP,PP1}.

In ref.~\cite{PP} we applied the sum rules technique for
the system of $\Upsilon$ resonances to determine the values of
the bottom quark pole mass $m_b$ and strong coupling constant 
$\al_s$. The analysis is based on the result for the heavy quark 
polarization function near the threshold in the 
next-to-next-to-leading order of perturbative 
QCD and relativistic expansion.
This result depends, in particular, on the two-loop correction
to the static potential of the quark-antiquark interaction
first computed in \cite{Peter}. Recently this correction 
has been recalculated independently with another technique
in \cite{YS}.
A different value of the coefficient $a_2$ came out
\[
a_2= \left({4343\over 162}+4\pi^2-{\pi^4\over 4}
+{22\over3}\zeta(3)\right)C_A^2-
\left({1798\over 81} + {56\over 3}\zeta(3)\right)C_AT_Fn_f
\]
$$
-\left({55\over 3} - 16\zeta(3)\right)C_FT_Fn_f
+\left({20\over 9}T_Fn_f\right)^2
$$
which is smaller than the previous result of ref.
\cite{Peter} by an amount $2\pi^2C_A^2$. 
After performing the analysis with the corrected 
value we found that this variation of the coefficient
affects our numerical estimates only slightly.
Namely, the value of $\al_s$ extracted from the sum
rules is practically insensitive to the above variation
while the value of $m_b$ decreases for $\sim 0.1\%$
when the corrected two-loop coefficient is used instead
of the previous one. Since the theoretical uncertainty 
in $m_b$ exceeds $1\%$ this variation is
negligible.

In the paper \cite{PP1} we used the relation 
between the moments of the $\Upsilon$ system spectral
density and the inclusive $B$-meson semileptonic width
for precise determination of the $|V_{cb}|$ matrix element.
Evaluating the moments we used
the next-to-next-to-leading 
order expression \cite{PP} of the heavy quark polarization function near the
threshold computed with Peter's coefficient $a_2$.
Changing it to the correct
Schr{\"o}der's value we obtain $\sim 0.1\%$
increase of the extracted numerical value for $|V_{cb}|$. 
The total theoretical uncertainty of this quantity 
exceeds $3\%$ that makes this variation completely
negligible.

The reason for such a weak influence of the change on our results is
that the coefficient $a_2$ parameterizes only a part of the correction to the
NRQCD Hamiltonian in this order  that makes the dependence 
on $a_2$ much softer than one could expect from the direct 
numerical change of the coefficient itself.  

To conclude, the numerical estimates of the bottom quark 
pole mass, the strong coupling constant and the
Cabibbo-Kobayashi-Maskawa matrix element $|V_{cb}|$ 
presented in refs~\cite{PP,PP1} 
are insensitive to the correction \cite{YS} of the previously
obtained value of the two-loop coefficient $a_2$ \cite{Peter}.
The corresponding shifts of the extracted values of $m_b$, 
$\al_s$ and $|V_{cb}|$ are an order of magnitude smaller than
the theoretical uncertainties of these quantities
given in \cite{PP,PP1}.

\end{document}